\documentclass{ws-ijprai}
\def\ArtDir{main/}
\begin{document}

%
\catchline{}{}{}{}{}
%
\title{Cellular Automata as a Network Topology}

\author{Temitayo Adefemi}

\address{University of Edinburgh\\
T.M.Adefemi@sms.ed.ac.uk}

\maketitle


\begin{abstract}
Cellular automata represent physical systems where both space and time are discrete, and the associated physical quantities assume a limited set of values. While previous research has applied cellular automata in modeling chemical, biological, and physical systems, its potential for modeling topological systems, specifically network topologies, remains underexplored. This paper investigates the use of cellular automata to model decentralized network topologies, which could enhance load balancing, fault tolerance, scalability, and the propagation and dissemination of information in distributed systems.
\end{abstract}


\vspace{6pt}

\section{Introduction}

Cellular automata (CA) are dynamic systems characterized by discrete space and time. Each cell in a regular lattice updates synchronously based on a deterministic rule. All cells follow the same rule and have a finite number of states.

Since their inception in the 1940s by John von Neumann and Stanislav Ulam, cellular automata have been reinvented multiple times under various names. Stephen Wolfram's extensive research has further highlighted the importance of CA in modeling complex systems and self-replicating mechanisms. A notable feature of cellular automata is their universality, meaning they can simulate any computation achievable by a computer algorithm, given the appropriate initial configuration and rules.

One prominent example is Conway's Game of Life, a universal cellular automaton capable of performing computations analogous to conventional computer algorithms. This example underscores CA's profound capabilities, which remain largely untapped in contemporary research.

The concept of cellular automata originates from von Neumann's idea of using computers to represent cells in the automaton. His initial design for a self-reproducing cellular automaton required each cell to support 29 states, and the array needed approximately 200,000 cells. However, subsequent research has expanded beyond von Neumann's original framework.

This paper reaffirms concepts that build on previous research. It applies cellular automata to model decentralized fault tolerance and resilience in network topologies. These models can enhance load balancing, information dissemination, propagation, and scalability in distributed systems.

We will illustrate these concepts by analogizing each system component to a cell within a lattice, following Von Neumann's foundational idea, where cell interactions mimic network dynamics. We will identify algorithms that enable information to spread through each systematic cell in the lattice and explore methodologies for paralleling the lattice. This allows one grid in the lattice to remain functional while others stay idle, which is beneficial for load balancing and parallelized networking. Additionally, we will introduce data scaling attributes within the lattice and other practical tools used in creating network topologies.

\section{Related Work}

The application of cellular automata (CA) in network topologies, load balancing, information dissemination, and scalability in distributed systems has been limited. Beerbohm et al. (1994) introduced a load-balancing strategy for distributed and parallel systems using cellular automaton systems to propose process migrations based on node load states, achieving significant performance improvements and scalability (Beerbohm et al., 1994). Similarly, Hofestädt et al. (1996) developed a cellular automaton system for simulating load balancing, demonstrating significant speedup and applicability in workstation clusters (Hofestädt et al., 1996). Shen and Zhu (2012) proposed the Cellular Automata Programming Algorithm (CAPA) for dynamic load balancing in large heterogeneous systems, showing excellent performance due to the simultaneity and concurrency of CA (Shen \& Zhu, 2012). In information dissemination, Luo (2022) analyzed data modeling for micro-propagation using CA, establishing models for network information dissemination coupled with data assimilation algorithms (Luo, 2022). Huang et al. (2019) proposed a CA-based propagation control mechanism to inhibit and monitor emergent-event contagion in sensor networks, enhancing invulnerability and scalability (Huang et al., 2019). In parallel computing and scalability, Mazzariol et al. (2000) presented an algorithm for dynamically remapping cells to balance load in parallel cellular automata applications, showing improved performance in load-unbalanced scenarios (Mazzariol et al., 2000), while Giordano et al. (2021) proposed an optimal workload assignment algorithm for parallel execution of CA, demonstrating significant performance improvements and reduction in execution times (Giordano et al., 2021). Furthermore, Souihi and Mellouk (2011) developed a knowledge dissemination mechanism for autonomic networks using CA to maintain knowledge planes, improving load balancing and scalability (Souihi \& Mellouk, 2011), and Cannataro et al. (1995) described CAMEL, a scalable CA environment for system modeling on parallel computers, achieving high performance in urban simulation applications (Cannataro et al., 1995). These studies highlight the versatility and effectiveness of cellular automata in addressing various challenges in distributed systems, from load balancing and information dissemination to scalability and fault tolerance.

\section{Cellular Automaton Applications in Network Topologies}

\subsection{Resillience \& Healing Mechanisms}
Network topologies are crucial for maintaining the robustness and reliability of communication systems. They are designed to be resilient, capable of managing failures, and able to prevent the escalation of errors. One innovative approach to enhancing the resilience of these systems is the use of cellular automata for network topology design.

Cellular automata are dynamic, discrete systems characterized by a grid of cells, each in one of a finite number of states, such as on or off. The state of each cell changes over time according to a set of rules that depend on the states of neighboring cells. This method is intriguing for network topology because each cell in the automaton can be conceptualized as a node in the network, such as a router or switch.

The essential advantage of using cellular automata in network topologies lies in their inherent reversibility. In many cellular automata systems, reversing any action or update is possible—effectively "undoing" the last movement or change. This feature can be precious in network management, particularly in error correction and system recovery scenarios. For example, if a node fails or a data transmission error occurs, the system can revert to a previous state before the failure, thereby providing a self-healing capability.

Furthermore, the local interaction rules in cellular automata can lead to complex global behaviors, making these systems capable of dynamically adapting to changes and failures. This adaptability can be tailored to mimic the behavior of network protocols that handle data transmission and error management, offering a more flexible and robust approach to designing network topologies.

Incorporating cellular automata into network design could lead to new types of network architectures that are more resilient to disruptions and capable of self-repair, significantly enhancing the reliability of data transmission networks.

\vspace{8pt}
Cellular automata (CA) provide a compelling model for decentralized computing systems due to their inherent design and operational characteristics. In a cellular automaton, each cell in the grid operates based on a set of rules that depend only on the state of its neighboring cells. This localized and independent decision-making process ensures the system is inherently decentralized, with no single cell acting as a central control point. This structure significantly enhances the resilience and fault tolerance of the network.

The decentralized nature of cellular automata means there is no single point of failure. In conventional centralized systems, the failure of a central unit can cripple the entire network. However, in a CA-based system, each cell functions independently. If one or more cells fail, the impact is generally localized, and neighboring cells can continue to operate normally, potentially adapting their behavior according to the local rules to compensate for the failure. This makes the network much more robust and resistant to failures.

Additionally, the ability of each cell to interact independently allows for greater adaptability and flexibility. In dynamic environments where network conditions or operational requirements change frequently, each cell's decentralized and independent operation enables the network to adapt more efficiently. Each cell can modify its behavior based on its immediate surroundings without awaiting instructions from a central authority, allowing the network to respond to changes in the environment quickly.

This model of operation, inspired by cellular automata, could be applied to various distributed computing applications, such as distributed data storage, peer-to-peer networks, and decentralized digital ledgers like blockchain. These applications can benefit significantly from the resilient, adaptable, and failure-resistant characteristics the cellular automaton approach provides.

\subsection{Load Balancing Mechanisms}.
Load balancing across distributed systems can be innovatively approached using a cellular automaton lattice. By imagining each cell within the automaton as a separate computer, this model allows for the decentralized and autonomous distribution of computational tasks. Simple yet effective rules can be employed to manage how computation is spread across this lattice, mimicking the structure and behavior of cellular automata.

For instance, cellular automaton rules could be designed to replicate various traditional load-balancing algorithms. A round-robin scheduling algorithm could be emulated by cyclically assigning tasks to each 'cell' or computer in a fixed order, ensuring that every cell handles an equal amount of load over time. Alternatively, a zigzag pattern could be utilized to distribute tasks in a way that dynamically adjusts based on the workload intensity at different nodes, enhancing efficiency and reducing bottlenecks.

Moreover, other patterns, such as weighted distribution or dynamic feedback-based approaches, could also be implemented. These methods would allow the system to adapt to changing load conditions in real-time, further optimizing computational efficiency and resource utilization across the network.

By leveraging the inherent properties of cellular automata—such as local interactions and simple rule-based evolution—this approach can lead to a robust, scalable, and flexible load-balancing strategy well-suited for complex distributed systems.

\subsubsection{Rules to emulate load balancing in a Cellular Automaton (CA)}

\paragraph{Neighbor activation rule}
In a cellular automaton modeled as a lattice, computation distribution can be cyclically managed across a grid of nodes, starting from an initial position and moving to the nearest unactivated neighbor sequentially. The process is defined mathematically as follows:

\vspace{9pt}
\textbf{Definitions:}
\begin{itemize}
    \item Let $C_{i,j}(t)$ represent the state of the node at position $(i,j)$ at time $t$, where $C_{i,j}(t) = 1$ indicates active, and $C_{i,j}(t) = 0$ indicates inactive.
    \item The lattice dimensions are $M \times N$.
\end{itemize}

\textbf{Initial Condition:}
\begin{equation}
    C_{0,0}(0) = 1, \quad C_{i,j}(0) = 0 \quad \text{for all} \quad (i, j) \neq (0, 0)
\end{equation}

\textbf{Activation Rule:}
\begin{equation}
    C_{k,l}(t) = \begin{cases} 
    1 & \text{if } (k, l) \text{ is the closest unactivated neighbor of } (i, j) \text{ where } C_{i,j}(t-1) = 1\\
    C_{k,l}(t-1) & \text{otherwise}
    \end{cases}
\end{equation}

\textbf{Restart and Cyclic Behavior:}
\begin{equation}
    C_{0,0}(t+1) = 1 \quad \text{if } \sum_{i,j} C_{i,j}(t) = M \times N
\end{equation}
\begin{equation}
    C_{i,j}(t+1) = 0 \quad \text{for all } (i, j) \neq (0, 0) \text{ if the above condition is met}
\end{equation}

\paragraph{Boundary feedback rule}
A boundary feedback rule can be effectively used to model network topologies within a cellular automaton and to emulate a round-robin distribution of computation. This rule can be implemented in a wrap-around format, where the computation, upon reaching the edge of the automaton, wraps around to the next row. This mechanism ensures that computation is smoothly distributed across the entire grid without interruption. Once the computation reaches the final cell of the automaton, it wraps back to the first cell, following a cyclic pattern similar to that used in the neighbor activation rule. This method ensures a continuous and evenly distributed computational load across the lattice, facilitating efficient processing in scenarios that mimic real-world network operations.

\vspace{9pt}

\textbf{Definitions:}
\begin{itemize}
    \item Let $C_{i,j}(t)$ represent the state of the node at position $(i,j)$ at time $t$, where $C_{i,j}(t) = 1$ indicates active, and $C_{i,j}(t) = 0$ indicates inactive.
    \item The lattice dimensions are $M \times N$.
\end{itemize}

\textbf{Initial Condition:}
\begin{equation}
    C_{0,0}(0) = 1, \quad C_{i,j}(0) = 0 \quad \text{for all} \quad (i, j) \neq (0, 0)
\end{equation}

\textbf{Boundary Feedback Activation Rule:}
\begin{equation}
    C_{k,l}(t+1) = \begin{cases} 
    1 & \text{if } (k, l) \text{ follows } (i, j) \text{ cyclically and } C_{i,j}(t) = 1\\
    0 & \text{otherwise}
    \end{cases}
\end{equation}
\begin{itemize}
    \item Here, $(k, l)$ is defined such that when $(i, j) = (m, n)$ reaches the boundary (either $m = M$ or $n = N$), it wraps around to the beginning of the next row or the start of the lattice.
\end{itemize}

\textbf{Restart and Cyclic Behavior:}
\begin{equation}
    C_{0,0}(t+1) = 1 \quad \text{if } \sum_{i,j} C_{i,j}(t) = M \times N
\end{equation}
\begin{equation}
    C_{i,j}(t+1) = 0 \quad \text{for all } (i, j) \neq (0, 0) \text{ if the above condition is met}
\end{equation}

\paragraph{Shift register rule}
A shift register rule can also be used to also model load balancing mechanisms, each cell in the grid has a state, which can change over time according to certain rules based on the states of its neighboring cells. To implement a round-robin system using a CA model, you would define a set of states—such as "active" and "inactive"—to represent the status of each cell at any given time. The round-robin approach can be modeled by cyclically shifting the "active" state among the cells in a predefined sequence, mimicking the scheduling technique used in computing where each process is given a fixed time slot, one after the other, in a cyclic order.

\vspace{9pt}

\textbf{Definitions:}
\begin{itemize}
    \item Let $C_{i,j}(t)$ represent the state of the cell at position $(i,j)$ at time $t$, where $C_{i,j}(t) = 1$ indicates the cell is active, and $C_{i,j}(t) = 0$ indicates the cell is inactive.
    \item The grid dimensions are $M \times N$, representing a matrix of cells.
\end{itemize}

\textbf{Initial Condition:}
\begin{equation}
    C_{0,0}(0) = 1, \quad C_{i,j}(0) = 0 \quad \text{for all} \quad (i, j) \neq (0, 0)
\end{equation}

\textbf{Shift Register Transition Rule:}
\begin{equation}
    C_{i,j}(t+1) = C_{i,(j-1) \mod N}(t) \quad \text{for all } i \text{ and } j
\end{equation}
This rule applies the shift register logic, cycling the active state through each column in a row and wrapping around the row boundaries.

\vspace{12pt}
\textbf{Boundary Feedback Activation Rule:}
\begin{equation}
    C_{i,0}(t+1) = C_{(i-1) \mod M,N-1}(t) \quad \text{for all } i
\end{equation}
This part of the rule ensures that the activation wraps around to the beginning of the next row when it reaches the end of the current row.

\vspace{12pt}
\textbf{Restart and Cyclic Behavior:}
\begin{equation}
    C_{0,0}(t+1) = 1 \quad \text{if } \sum_{i,j} C_{i,j}(t) = M \times N
\end{equation}
\begin{equation}
    C_{i,j}(t+1) = 0 \quad \text{for all } (i, j) \neq (0, 0) \text{ if the above condition is met}
\end{equation}

This part governs the restart mechanism, ensuring that once all cells have been activated in turn, the entire grid resets to start the round-robin process anew.

\vspace{12pt}
These simple rules can be manipulated to not only emulate round robin strategies but to also create multiple formats of load balancing algorithms. Neighbor activation rule can effectively be changed to zigzag neighbour activation rule in which we activate the neighbor in a zigzag pattern, this can also be implemented using the other rules such as shift register rule

\subsection{Information Propagation \& Dissemination}

In cellular automata, information spreads across the lattice through various rules that utilize the model's inherent characteristics. A critical method involves applying load balancing rules, which help distribute computational tasks and data evenly across the network. This prevents any single node from becoming overloaded, enhancing efficiency and reducing latency. Specifically, the neighbor activation rule allows for efficient information propagation. This rule is beneficial for performing heartbeat checks within a network topology. For example, when a byte of data is sent to a neighboring computer in the lattice, a response from that computer confirms its active status. Without a response, mechanisms could be implemented to restart the system. This demonstrates how cellular automata can effectively manage and optimize network topologies, facilitating reliable and efficient information transfer between systems in the lattice. Here are rules which could be used to illustrate the methodology.

\paragraph{Majority Rule}
One effective rule for information dissemination is the majority rule, where a cell adopts the state that is most common among its neighbors. This rule helps in reaching consensus and propagating dominant states across the automaton.

\paragraph{Rule 110 and Universal Computation}
Rule 110 is an elementary cellular automaton rule known for its ability to perform universal computation, meaning it can simulate any Turing machine. This rule is notable for its complex behavior and ability to propagate information effectively.

\paragraph{Genetic Algorithms for Rule Optimization}
Genetic algorithms have been used to evolve cellular automata rules for specific tasks, such as the majority classification problem. These algorithms optimize the rules to improve the dissemination and integration of information across the cellular space.

\vspace{12pt}

\textbf{Definitions:}
\begin{itemize}
    \item Let $C_{i,j}(t)$ represent the state of the cell at position $(i,j)$ at time $t$, where $C_{i,j}(t) = 1$ indicates the cell is active, and $C_{i,j}(t) = 0$ indicates it is inactive.
    \item The lattice dimensions are $M \times N$.
\end{itemize}

\textbf{Initial Condition:}
\begin{equation}
    C_{0,0}(0) = 1, \quad C_{i,j}(0) = 0 \quad \text{for all} \quad (i, j) \neq (0, 0)
\end{equation}

\textbf{Activation Rule (Majority Rule):}
\begin{equation}
    C_{i,j}(t+1) = 
    \begin{cases} 
    1 & \text{if the majority of the neighbors of } (i, j) \text{ at time } t \text{ are 1} \\
    0 & \text{otherwise}
    \end{cases}
\end{equation}

\textbf{Rule 110:}
\begin{equation}
    C_{i,j}(t+1) = C_{i-1,j}(t) \oplus (C_{i,j}(t) \lor C_{i+1,j}(t))
\end{equation}
Here, $\oplus$ represents the XOR operation, and $\lor$ represents the OR operation.

\vspace{10pt}
\textbf{Genetic Algorithms for Rule Optimization:}
Optimization using genetic algorithms for tasks like the majority classification problem involves iteratively adjusting the rules based on fitness, which measures how well the rules achieve desired information propagation.

\vspace{10pt}
\textbf{Restart and Cyclic Behavior:}
\begin{equation}
    C_{0,0}(t+1) = 1 \quad \text{if } \sum_{i,j} C_{i,j}(t) = M \times N
\end{equation}
\begin{equation}
    C_{i,j}(t+1) = 0 \quad \text{for all } (i, j) \neq (0, 0) \text{ if the above condition is met}
\end{equation}

\subsection{Parallelization \& Concurrency}

A cellular automaton, defined on an N×M lattice, is a practical framework for implementing parallelization and concurrency in computational environments. This lattice structure inherently supports simultaneous operations across its multiple rows and columns. The parallel processing capability is governed by predefined rules, which dictate the initiation and progression of computations across the lattice.

For instance, one such rule could be simultaneously initiating computation at the start of each row or column. This approach leverages the discrete nature of the cellular automaton, where each cell's state is updated based on its neighbors, allowing for independent updates within each row or column. As computations proceed, they can advance concurrently, effectively distributing the processing load and reducing overall computation time.

Once the computation reaches the end of a row or column, it can be designed to either wrap around to the beginning of the same row or column or terminate, depending on the model's specific requirements. This flexibility in defining boundary conditions and computational rules enhances the adaptability of cellular automata in various applications, ranging from simple data processing tasks to complex simulations of physical systems.

Moreover, the parallelization enabled by cellular automata is not merely limited to linear advancements across rows or columns. It can also be extended to more complex patterns of computational progression, such as diagonal processing or expansion in multiple directions. It can benefit simulations requiring more dynamic interaction between cells across the lattice.

In summary, a cellular automaton in an N×M lattice format offers a robust method for achieving efficient parallel computations. This method optimizes computational speed and efficiency and provides a versatile tool for exploring various scientific and engineering problems where concurrency and parallel processing are crucial.

\section{Conclusion}
The paper explores the application of cellular automata (CA) in decentralized network topologies, highlighting their potential to enhance resilience, load balancing, and information dissemination in distributed systems. Cellular automata's inherent properties, such as localized interaction, reversibility, and decentralized control, make them suitable for designing robust and adaptable network architectures. These features allow for efficient load balancing and fault tolerance, as well as dynamic adaptation to changing conditions, which are crucial for maintaining the reliability and efficiency of communication networks. The study reaffirms the versatility and effectiveness of CA in addressing various challenges in distributed systems, from load balancing and information dissemination to scalability and fault tolerance.

\end{document}